\documentclass[12pt,twoside]{article}
\usepackage{amsmath,amsfonts,amssymb}
\usepackage{latexsym}
\usepackage{dcolumn}
\usepackage{graphicx,epsfig}
\usepackage{amsthm}
\usepackage{hyperref}

\begin{document}

\begin{flushright}
{\tt IISER(Kolkata)/GR-QC\\ \today}
\end{flushright}
\vspace{1.2cm}

\begin{center}
{\Large \bf The Perihelion Precession of Mercury and the Generalized Uncertainty Principle }
\vglue 0.5cm
Barun Majumder\footnote{barunbasanta@iiserkol.ac.in}
\vglue 0.6cm
{\small {\it Department of Physical Sciences,\\Indian Institute of Science Education and Research (Kolkata),\\
Mohanpur, Nadia, West Bengal, Pin 741252~,\\India}}
\end{center}
\vspace{.1cm}

\begin{abstract} 
\hspace{-5pt} Very recently authors in \cite{d5} proposed a new Generalized Uncertainty Principle (or GUP) with a linear term in Plank length. In this Letter the effect of this linear term is studied perturbatively in the context of Keplerian orbits. The angle by which the perihelion of the orbit revolves over a complete orbital cycle is computed. The result is applied in the context of the precession of the perihelion of Mercury. As a consequence we get a lower bound of the new intermediate length scale offered by the GUP which is approximately $40$ orders of magnitude below Plank length.
\vspace{5mm}\newline Keywords: GUP, minimum length, Hamilton vector, Mercury perihelion
\end{abstract}


The idea that the uncertainty principle could be affected by gravity was first given by Mead \cite{d1}. Later modified commutation relations between position and momenta commonly
known as Generalized Uncertainty Principle ( or GUP ) were given by candidate theories of quantum gravity ( String Theory, Doubly Special
Relativity ( or DSR ) Theory and Black Hole Physics ) with the prediction of a minimum measurable length \cite{d2,c4,d3}. Similar kind of
commutation relation can also be found in the context of Polymer Quantization in terms of Polymer Mass Scale \cite{d4}.
\par
The authors in \cite{d5} proposed a GUP which is consistent with DSR theory, String theory and Black Hole Physics and which says
\begin{equation}
\left[x_i,x_j\right] = \left[p_i,p_j\right] = 0 ,
\end{equation}
\begin{equation}
\label{e2}
[x_i, p_j] = i \hbar \left[  \delta_{ij} -  l  \left( p \delta_{ij} +
\frac{p_i p_j}{p} \right) + l^2  \left( p^2 \delta_{ij}  + 3 p_{i} p_{j} \right)  \right],
\end{equation}
\begin{align}
 \Delta x \Delta p &\geq \frac{\hbar}{2} \left[ 1 - 2 l <p> + 4 l^2 <p^2> \right]  \nonumber \\
& \geq \frac{\hbar}{2} \left[ 1  +  \left(\frac{l}{\sqrt{\langle p^2 \rangle}} + 4 l^2  \right)  \Delta p^2  +  4 l^2 \langle p \rangle^2 -  2 l \sqrt{\langle p^2 \rangle} \right],
\end{align}
where $ l=\frac{l_0 l_{pl}}{\hbar} $. Here $ l_{pl} $ is the Plank length ($ \approx 10^{-35} m $). It is normally assumed that the dimensionless
parameter $l_0$ is of the order unity. If this is the case then the $l$ dependent terms are only important at or near the Plank
regime. But here we expect the existence of a new intermediate physical length scale of the order of $l \hbar = l_0 l_{pl}$. We also note
that this unobserved length scale cannot exceed the electroweak length scale \cite{d5} which implies $l_0 \leq 10^{17}$. These equations are
approximately covariant under DSR transformations but not Lorentz covariant \cite{d3}. These equations also imply
\begin{equation}
\Delta x \geq \left(\Delta x \right)_{min} \approx l_0\,l_{pl}
\end{equation}
and
\begin{equation}
\Delta p \leq \left(\Delta p \right)_{max} \approx \frac{M_{pl}c}{l_0}
\end{equation}
where $ M_{pl} $ is the Plank mass and $c$ is the velocity of light in vacuum. It can be shown that equation (\ref{e2}) is satisfied by the
following definitions $x_i=x_{oi}$ and $p_i=p_{oi} (1 - l\,p_o + 2\,l^2\,p_o^2)$, where $x_{oi}$, $p_{oj}$ satisfies $[x_{oi}, p_{oj}]= i \hbar \delta_{ij}$. Here we can interpret $p_{oi}$ as the momentum at low energies having the standard representation in position space ($ p_{oi}=-i\hbar \frac{\partial}{\partial x_{oi}}$) with $p_o^2=\sum_{i=1}^3 p_{oi}p_{oi}$ and $ p_i $ as the momentum at high energies. It can also be shown that any non-relativistic Hamiltonian of the form ${\cal H}=\frac{p^2}{2m} + V(r)$ can be written in the form \cite{d5}
\begin{equation}
\label{e7}
 {\cal H}=\frac{p_o^2}{2m} + V(r) - \frac{l}{m}p_o^3 + {\cal O}(l^2)+\ldots.
\end{equation}
Here we neglect terms $ {\cal O}(l^2)$ and higher in comparison to terms $ {\cal O}(l)$ to study the effect of the linear term in $l$ in the first approximation as $l=l_0\,l_{pl}$. The effect of this proposed GUP is well studied for some well known quantum mechanical Hamiltonians in \cite{d5,d6} and also studied in the context of quantum cosmological models \cite{d}, Friedmann equations and black hole evaporation \cite{zz}.
\par
In this Letter we will study the effect of the third term ( which is linear in Plank length ) perturbatively in the context of Kepler orbits. To start with we would like to comment
that besides the Laplace–Runge–Lenz vector \cite{b38} there is also another conserved vector quantity which is known as the Hamilton vector. Due to the lack of evidence we can only cite few
documents \cite{b39} from where we can get precise knowledge about this Hamilton vector. In the presence of perturbation the perihelion of the orbit begins to
precess. And the precession rate of the Hamilton vector coincides with the precession rate of the perihelion \cite{c7}. In the presence of perturbation we can calculate the precession rate
of the Hamilton vector and hence we can compute the angle by which the perihelion revolves over a complete orbital cycle.\\
The Hamilton vector is written as 
\begin{equation}
\label{e8}
  \vec{{\bf u}} = \frac{\vec{p}}{m} - \frac{k}{L}\,\hat{\phi}, 
\end{equation}
where $k=GmM$ and $L=mr^2{\dot \phi}$ is the orbital momentum. The unit vector
\begin{equation}
\label{e9}
 \hat{\phi} = \frac{\vec{L}\times \vec{r}}{Lr}
\end{equation}
lie on the plane of the orbit. For the Keplerian orbits we write equation (\ref{e7}) as 
\begin{equation}
 {\cal H} = \frac{p^2}{2m} - \frac{k}{r} - \frac{l}{m}\,p^3 = {\cal H}_0 - \frac{l}{m}\,p^3 ,
\end{equation}
where ${\cal H}_0 = \frac{p^2}{2m} - \frac{GmM}{r}$ is the unperturbed Hamiltonian. As mentioned before we will treat the term which is linear in Plank length as the perturbation. Now we will
follow the method introduced in \cite{c} for computing the precession rate of the perihelion of the orbit. The Hamilton vector is no longer conserved in the presence of perturbation and we can
write
\begin{equation}
 \dot{\vec{{\bf u}}} = \{\vec{{\bf u}},{\cal H}\} = -\frac{l}{m}\,\{\vec{{\bf u}},p^3\} = -\frac{3\,l\,p^2}{m}\,\{\vec{{\bf u}},p\}.
\end{equation}
A straightforward calculation using equations (\ref{e8}) and (\ref{e9}) yields
\begin{align}
\label{e12}
 \dot{\vec{{\bf u}}} &= -\frac{3\,l\,k\,p}{m\,r^3\,L^2}\,\vec{L}\times(\vec{r}\times\vec{L}) \nonumber \\
&= -\frac{3\,l\,k\,p}{m\,r^3}\,\vec{r}.
\end{align}
Equation $\{2.3\}$ of \cite{c9} gives the precession rate of $\vec{{\bf u}}$ as
\begin{equation}
\label{e13}
 \vec{\omega} = \frac{\vec{{\bf u}}\times \dot{\vec{{\bf u}}}}{{\bf u}^2},
\end{equation}
 where ${\bf u}=\frac{k\,e}{L}$ and $e$ is the eccentricity of the elliptic orbit and $0<e<1$. Using equations (\ref{e8}),(\ref{e9}),(\ref{e12}) and (\ref{e13}) we finally get
\begin{equation}
\label{e14}
 \omega = \frac{3\,l\,p\,L}{m\,r^2\,e^2}\bigg(\frac{R}{r}-1\bigg),
\end{equation}
where $R=\frac{L^2}{km}$ is the semi-latus rectum of the unperturbed orbit. In the first order of approximation we will use the known relations
\begin{equation}
\label{e15}
 \frac{R}{r} = 1 + e\,\mbox{cos} \phi
\end{equation}
and
\begin{equation}
\label{e16}
 \frac{p^2}{2m} - \frac{k}{r} = -\frac{k}{2a},
\end{equation}
where $a=\frac{R}{1-e^2}$ is the semi-major axis of the elliptic orbit. As the precession rate of the Hamilton vector equals the precession rate of the perihelion \cite{c7}, so the angle
by which the perihelion revolves over a complete orbital cycle due to the presence of the perturbation can be written as
\begin{equation}
 \Delta \theta = \int_0^T \omega\, dt = \int_0^{2\pi}\, \frac{\omega}{\dot \phi} \,d\phi.
\end{equation}
 With the use of equations (\ref{e14}),(\ref{e15}) and (\ref{e16}) we can finally get
\begin{align}
\label{e18}
 \Delta \theta &= 3\,l\,\sqrt{\frac{mk}{R}}\,\frac{1}{e}\,\int_0^{2\pi}\,\mbox{cos}\,\phi\,[1+e^2+2e\,\mbox{cos}\,\phi]^{\frac{1}{2}}\,d\phi \nonumber \\
&= 3\,l\,\sqrt{\frac{mk}{R}} \, F,
\end{align}
where $F = \frac{1}{e}\,\int_0^{2\pi}\,\mbox{cos}\,\phi\,[1+e^2+2e\,\mbox{cos}\,\phi]^{\frac{1}{2}}\,d\phi$ and the integral can be easily evaluated numerically for a particular value of $e$. Now with $k=GmM$ and $l=\frac{l_0}{M_{pl}c}=\frac{l_0l_{pl}}{\hbar}$ we get equation (\ref{e18}) in the form
\begin{equation}
\label{e19}
 \Delta \theta = 3\,l_0\,\frac{m}{M_{pl}}\,\sqrt{\frac{(2GM/c^2)}{(2R)}}\,\,F\,\,.
\end{equation}
For the planet Mercury, the parameters which govern the motion are \cite{sandor18}
\begin{align}
\label{mer} 
\bigg\{\,\,
\frac{2GM_{\odot}}{c^2} &= 2.95325008 \times 10^3 \,\,\text{metre}\,, \nonumber \\
m &= 3.3022 \times 10^{23} \,\,\text{kg} \,, \nonumber \\
a &= 5.7909175 \times 10^{10}\,\, \text{metre} \,, \nonumber \\
e &= 0.20563069\,.\,\, \bigg\}
\end{align}
Here $M_{\odot}$ is the solar mass and $a$ is the semi-major axis of the elliptic orbit and $R=a(1-e^2)$ is the semi-latus rectum of the orbit. The observed
value of $\Delta \theta$ for the precession of the perihelion of Mercury is \cite{sandor17}
\begin{equation}
\Delta \theta_{obs} = 2\,\pi\,(7.98734 \pm 0.00037) \times 10^{-8}\,\,\text{radian/revolution}\,\,. 
\end{equation}
Einstein's general relativity explains this observation very accurately. Following \cite{c1} and references therein we can see
\begin{equation}
\label{dif}
\Delta \theta_{obs} - \Delta \theta_{GR} = 2\,\pi\,(-0.00010 \pm 0.00037) \times 10^{-8} \,\,\text{radian/revolution}\,\,.
\end{equation}
We may feel that there is nothing left to explain but still we can also see that the difference is not exactly zero. Using (\ref{mer}) we can evaluate
$\Delta \theta$ from equation (\ref{e19}) and we get
\begin{equation}
\label{de1}
\Delta \theta = 2.3209 \times 10^{28} \,\, l_0  \,\, \text{radian/revolution}\,\,.
\end{equation}
Believing that the remnant difference of (\ref{dif}) is explained by (\ref{de1}) we get a lower bound of the new intermediate physical length
scale ($\approx l_0\,l_{pl}$) with
\begin{equation}
l_0 (\text{min.}) = 7.31 \times 10^{-40}\,\,.
\end{equation} 
As an interesting fact we cannot neglect the existence of a new length scale $\sim 10^{-75}$ metre.    
\par
So in this Letter we have considered the recently proposed GUP \cite{d5} which is approximately covariant under DSR transformation \cite{d3} but not Lorentz covariant. The GUP has a linear term in Plank length. We have studied the effect of that linear term in Plank length in the context of Keplerian orbits. For calculating the precession rate of the perihelion we have adapted the method introduced in \cite{c} and the whole approach is perturbative. The author in \cite{c} made use of the modified commutation relation of the form \cite{c4}
\begin{equation}
 [\hat x_i,\hat p_j] = i\,\hbar \,\, (\delta_{ij} + \beta \hat p^2 \delta_{ij} + \beta' \hat p_i \hat p_j ),
\end{equation}
where the Plank length is given by $l_{pl} \approx \hbar\, \sqrt{3\beta+\beta'}$. So the effect studied in \cite{c} is ${\cal O}(l_{pl}^2)$ and higher order terms are neglected perturbatively. There the results were applied to earth-bound satellites where the linear momentum of the satellite should be much less than $6.6$ kg m/s. This seems quite strange to us because all planetary motions are always accompanied by very high momenta.
\par
Here we have equated our result with the precession of perihelion of Mercury. Though the observed precession rate is well
explained by general relativity but up to order of $10^{-8}$. Our equation leads to a prediction of the existence of a new length scale which is $40$ orders
of magnitude below Plank length. More briefly we have succeeded to set a lower bound to the new intermediate length scale offered by the GUP introduced
in \cite{d5}. Similar result was also found earlier in \cite{c1} where the idea of non-commuting spatial co-ordinates was taken into account. It was shown that
the precession of the perihelion of Mercury restricts the value of the minimum length which is $33$ orders of magnitude lower than Plank length.  

\section*{Acknowledgements}
The author is very much thankful to Prof. Narayan Banerjee for helpful discussions and guidance.

\end{document}